\documentclass[showpacs,amssymb,twocolumn,aps,nofootinbib]{revtex4}
\usepackage{amsmath}
\usepackage{amstext}
\usepackage{amsopn}
\usepackage{amsfonts}
\usepackage{amssymb}
\usepackage{accents}
\usepackage{empheq}
\usepackage{graphicx}
\usepackage{epsf}
\usepackage{graphics}
\usepackage{xcolor}
\usepackage[latin1]{inputenc}

\allowdisplaybreaks[3]

\begin{document}

\title{Generalized Kadanoff-Baym relation in nonequilibrium quenched models}

\author{A. L. M. Britto$^{a}$, Ashok K. Das$^{b,c}$ and J. Frenkel$^{a}$}
\affiliation{$^{a}$ Instituto de Física, Universidade de São Paulo, 05508-090, São Paulo, SP, Brazil}
\affiliation{$^b$ Department of Physics and Astronomy, University of Rochester, Rochester, NY 14627-0171, USA}
\affiliation{$^c$ Saha Institute of Nuclear Physics, 1/AF Bidhannagar, Calcutta 700064, India}

\begin{abstract}
 In the context of a broad class of quenched models, we derive a generalized differential form 
of the Kadanoff-Baym (KB) ansatz which relates the out of equilibrium correlated and spectral Green's functions. This relation holds at any time both before  the quench (when it coincides with the fluctuation-dissipation theorem) as well as after it. We also examine, in the context of exactly soluble quenched models, the validity of some of the earlier alternative extensions of the KB ansatz.
\end{abstract}

\maketitle

\section{Introduction}
The fluctuation-dissipation theorem \cite{callen,kubo,frenkel} plays an important role in the study of systems in thermal equilibrium. This result expresses a general relation between the statistical fluctuations in the system and its response to weak external perturbations. The fluctuations and the response of the system may be described by the correlated and spectral Green's functions defined respectively (in a scalar field theory) as
\begin{align}
iG_c(x,y) & =\langle[\phi(x),\phi(y)]_+\rangle,\notag\\
iG_{\rho}(x,y) & =\langle[\phi(x),\phi(y)]\rangle,\label{1}
\end{align}
where the angular brackets represent thermal averages. The fluctuation-dissipation theorem can be written in terms of their Fourier transforms as ($k_\mu$ corresponds to the momentum conjugate to the coordinate difference $x^\mu -y^\mu$)
\begin{equation}
\widetilde{G}_{c} (k) = \left(1 + 2 N(k_{0})\right) \widetilde{G}_{\rho} (k),	\label{1a}
\end{equation}
where $N (k_{0}) = 1/(e^{\beta k_{0}} -1)$ denotes the equilibrium Bose-Einstein distribution function. Here one has used the fact that in equilibrium the Green's functions are time translation invariant, which does not hold when the system is out of equilibrium. The fluctuation-dissipation theorem has been generalized by Kadanoff and Baym to systems near equilibrium \cite{KB}. Their ansatz states that, when Fourier transformed, the two Green's functions satisfy near equilibrium the relation
\begin{equation}\widetilde{G}_c(k,T) = \left(1+2f(k_0,T)\right)\widetilde{G}_{\rho}(k,T),\label{2}
\end{equation}
where $T=(x^0+y^0)/2$ and $f(k_0,T)$ is an appropriate distribution function.
      
The study of out of equilibrium processes is of much interest in various branches of physics such as cosmology, high energy and condensed matter physics \cite{boya,aarts,berges,lindner,marinari,cristiani,altland}. Therefore, it is quite important to obtain consistent  generalizations of the Kadanoff-Baym  ansatz for these regimes. There have been several proposals in this respect both in the context of relativistic \cite{calzetta,millington,boya2} and non-relativistic quantum field theories \cite{daniele,lipavski,spicka}. Simple soluble 
models \cite{das,frenkel2} provide a testing ground for the validity of these proposals as well as for constructing further generalizations. In a previous paper \cite{britto} we have studied such an extension, which may describe the large time behaviour in glassy systems with an effective temperature, within the context of an exactly soluble quenched model. However, this approach is not suitable for studying the behavior of systems soon after the quench.
      
In this work we discuss a more general treatment which is applicable at any time, both before and after the quench. In section \ref{sec2}, we consider a broad class of nonequilibrium quenched models ( mostly not exactly soluble), which also includes some soluble models. { We}  point out several basic features of  the exact Green's functions following from the  results obtained in these soluble models. In section \ref{sec3}, we derive a generalized differential form of the Kadanoff-Baym ansatz for the whole class of quenched models, relating the exact out of equilibrium spectral and correlated Green's functions, which holds at all times. This generalized KB relation shows that various features of the exact Green's functions necessarily appear in generic quenched models, as a consequence of causality. In section \ref{sec4} we examine, within the context of non-relativistic soluble quenched models, some of the earlier extensions of the KB ansatz \cite{daniele,lipavski,spicka}. We find that these  proposals involve certain assumptions, concerning the behaviour of out of equilibrium Green's functions, which might be appropriate only in some time sectors. We conclude the paper in section \ref{sec5}, which presents a brief summary of the main results.

\section{Exact Green's functions in a quenched model}\label{sec2}

We study a simple out of equilibrium quantum field theory which describes a free real scalar field of mass $m$ at negative times, $x^{0} < 0$ (the reference time can be arbitrary, but for simplicity we choose it to be zero). An effective mass correction is introduced for $x^{0}\geq 0$ so that the Lagrangian density which describes the system is given by 
\begin{equation}
\mathcal{L}=\frac{1}{2}\partial_\mu\phi\partial^\mu\phi-\frac{m^2}{2}\phi^2-\frac{1}{2}\Pi(x^0)\phi^2,\label{3}
\end{equation}
where the interaction term due to the sudden quench  $\Pi(x^0)$ takes the theory out of equilibrium.
In particular, if we choose a special quench of the form
\begin{equation}
\Pi(x^0)=\delta m^2\,\theta(x^0)+\Delta m\,\delta(x^0),\label{4}
\end{equation}
then the model is exactly soluble when either  $\delta m^2$ or $\Delta m$ vanishes. In this section, we consider 
the model with  $ \Delta m=0$  which was studied in earlier papers and we discuss the model with $\delta m^2=0$
in the appendix. The general case, which involves a generic quench $\Pi(x^0)$ in \eqref{3},  will be  examined in the next section.

The exact retarded Green's function for the model
\begin{equation}
iG_R(x,y)=\theta(x^0-y^0)\langle[\phi(x),\phi(y)]\rangle,\label{5}
\end{equation}
and the exact correlated Green's function \eqref{1} have been calculated \cite{frenkel2} and yield the mixed space forms
\begin{widetext}
\begin{align}
G_R(x^0,y^0,\omega) & = \theta(x^0-y^0)\left[-\frac{1}{\omega}\theta(-x^0)\theta(-y^0)\sin \omega(x^0-y^0)-\frac{1}{\Omega}\theta(x^0)\theta(y^0)\sin \Omega(x^0-y^0)\right.\nonumber
\\
&\quad\left. +\frac{1}{2\omega}\theta(x^0)\theta(-y^0)\left(\left(1-\frac{\omega}{\Omega}\right)\sin(\Omega x^0+\omega y^0)-\left(1+\frac{\omega}{\Omega}\right)\sin(\Omega x^0-\omega y^0)\right)\right],\label{7}\\
G_c(x^0,y^0,\omega) & =\frac{\coth \frac{\beta\omega}{2}}{i\omega}\left[\frac{}{}\theta(-x^0)\theta(-y^0)\cos \omega(x^0-y^0)\right.\nonumber\\
&\quad +\frac{1}{2}\theta(x^0)\theta(-y^0)\left(\left(1+\frac{\omega}{\Omega}\right)\cos(\Omega x^0-\omega y^0)+\left(1-\frac{\omega}{\Omega}\right)\cos(\Omega x^0+\omega y^0)\right)\nonumber\\
&\quad +\frac{1}{2}\theta(-x^0)\theta(y^0)\left(\left(1+\frac{\omega}{\Omega}\right)\cos(\omega x^0-\Omega y^0)+\left(1-\frac{\omega}{\Omega}\right)\cos(\omega x^0+\Omega y^0) \right)\nonumber\\
&\quad  +\left.\frac{1}{2}\theta(x^0)\theta(y^0)\left(\left(\frac{\Omega^2+\omega^2}{\Omega^2}\right)\cos\Omega(x^0-y^0)+\left(\frac{\Omega^2-\omega^2}{\Omega^2}\right)\cos\Omega( x^0+y^0)\right)\right],\label{6}
\end{align}
\end{widetext}
where $\beta$ is the inverse of the initial equilibrium temperature  (in units of the Boltzmann constant) and we have defined
\begin{equation}
\omega^2=\mathbf{k}^2+m^2,\qquad\Omega^2=\omega^2+\delta m^2.\label{8}
\end{equation}

Equations \eqref{7}  and \eqref{6} can now be Fourier transformed with respect to the time difference $t=x^0-y^0$ 
\begin{equation}
\widetilde{G}_{c,\rho}(k_0,T,\omega)=\int_{-\infty}^{\infty} dt\, e^{ik_0 t}\,G_{c,\rho}(t,T,\omega),\label{10.2}
\end{equation}
where $T=\frac{x^{0}+y^{0}}{2}$ and lead to  
\begin{widetext}
\begin{align}
{\rm Im}\, \widetilde{G}_R(k_0,T,\omega) & =\frac{\widetilde{G}_\rho (k_0,T,\omega)}{2i}\nonumber\\
&=\frac{\theta(T)}{2\Omega}\left[\frac{\sin 2(k_0+\Omega)T}{k_0+\Omega}-\frac{\sin 2(k_0-\Omega)T}{k_0-\Omega}\right]
+\frac{\theta(-T)}{2\omega}\left[\frac{\sin 2(k_0-\omega)T}{k_0-\omega}-\frac{\sin 2(k_0+\omega)T}{k_0+\omega}\right]
\nonumber \\
&\quad +\left\{
\frac{\nu_+}{2\omega\Omega}\left[\theta(T)\left(\frac{\sin 2(k_0-\Omega)T}{k_0-\nu_+}-\frac{\sin 2(k_0+\Omega)T}{k_0+\nu_+}\right)\right.\right.
\nonumber \\
&\quad + \theta(-T)\left(\frac{\sin 2(k_0+\omega)T}{k_0+\nu_+}-\left.\left.\frac{\sin 2(k_0-\omega)T}{k_0-\nu_+}\right)\right]+(\omega\rightarrow-\omega)\right\}
\nonumber\\
&\quad +\pi\left\{\frac{\nu_+}{2\omega\Omega}\cos (2\nu_- T) \left[\delta\left(k_0+\nu_+\right)-\delta\left(k_0-\nu_+\right)\right]+(\omega\rightarrow-\omega)\right\},\label{10}
\end{align}
where we have defined
\begin{equation}
\nu_+=\frac{\Omega+\omega}{2}~~~;~~~\nu_-=\frac{\Omega-\omega}{2}
\end{equation}
so that $\nu_+\leftrightarrow\nu_-$ when $\omega\rightarrow-\omega$. Moreover, we obtain
\begin{align}
\widetilde{G}_c(k_0,T,\omega) & =\frac{\coth\frac{\beta\omega}{2}}{i\omega}
\left\{ -\theta(-T)\left[\frac{\sin 2(k_0+\omega)T}{k_0+\omega}+\frac{\sin 2(k_0-\omega)T}{k_0-\omega}\right]\right.\nonumber\\
&\quad + \theta(T)\left[\frac{\Omega^2+\omega^2}{2\Omega^2}\left(\frac{\sin 2(k_0+\Omega)T}{k_0+\Omega}+\frac{\sin 2(k_0-\Omega)T}{k_0-\Omega}\right)+
\frac{\Omega^2-\omega^2}{\Omega^2}\cos(2\Omega T)\frac{\sin 2k_0T}{k_0}
\right]\nonumber\\
&\quad + \left[-\theta(T)\frac{\nu_+}{\Omega}
\left(\frac{\sin 2(k_0+\Omega)T}{k_0+ \nu_+}+\frac{\sin 2(k_0-\Omega)T}{k_0-\nu_+}\right)
\right.+\nonumber\\
&\quad +
\left.\theta(-T)\frac{\nu_+}{\Omega}
\left(\frac{\sin 2(k_0+\omega)T}{k_0+\nu_+}+\frac{\sin 2(k_0-\omega)T}{k_0-\nu_+}\right)+(\omega\rightarrow -\omega)
\right]\nonumber\\
 &\quad +\pi
 \Big[
 \frac{\nu_+}{\Omega}\cos (2\nu_- T)
\left[
 \delta\left(k_0+\nu_+\right)+\delta\left(k_0-\nu_+\right)\right]+\left.(\omega\rightarrow -\omega)
 \Big]
 \right\},
 \label{9}
\end{align}
\end{widetext}
where we have used  the relations between the spectral, retarded and the advanced Green's functions, namely, 
\begin{align}
G_\rho(x,y) & = G_R(x,y)-G_A(x,y)\notag\\
 & = G_R(x,y)-G_R (y,x).\label{10.1}
\end{align}
(In the Fourier transformed space, the last form of the relation in \eqref{10.1} gives $\tilde{G}_\rho=2i\, \text{Im}\, \tilde{G}_{R}$.)

There are several interesting features to be noted here.  Both the functions in \eqref{10} and \eqref{9} 
have finite as well as pole terms in the $k_0$ space. The finite contributions come from the region in \eqref{10.2}  where the $t$-integration is bounded, namely, $|t|<2|T|$. From the relations $x^0=T+t/2, y^0=T-t/2$, it follows that this range corresponds to the case when $x^0$ and $y^0$ have the same sign. From the structures of \eqref{10} and \eqref{9}, it appears that such contributions to the Green's functions $\widetilde{G}_c$ and $\widetilde{G}_\rho$ may be related in a non-trivial way. On the other hand, the pole terms at $k_0=\nu_{\pm}$ arise from the region in \eqref{10.2} where the $t$-integration is unbounded, namely, $2|T|<|t|<\infty$ which correspond to the case where $x^0$ and $y^0$ have opposite signs. Since $\widetilde{G}_c(k_0)$ is an even functions of $k_0$ while $\widetilde{G}_\rho(k_0)$ is odd, the pole terms also appear at the frequencies $(-\nu_{\pm})$. One can see from \eqref{10} and \eqref{9} that the values of $\widetilde{G}_c(k_0,T,\omega)$ and $\widetilde{G}_\rho(k_0,T,\omega)$ near the poles at $k_{0}= \nu_{\pm}$  are related in a simple way (through the Bose-Einstein distribution)
\begin{equation}
\widetilde{G}_c(k_0,T,\omega)
 \simeq \pm\left[1+2 N\left(\frac{\beta\omega k_0}{\nu_{\pm}}\right)\right]
\widetilde{G}_\rho(k_0,T,\omega).\label{11}
\end{equation}
We note that the factor in the square bracket can also be written as $\coth[\frac{\beta\omega k_0}{2\nu_{\pm}}]$, which is an odd function of $k_0$
as required by the consistency of \eqref{11} under $k_0\rightarrow -k_0$.
We finally remark that the contributions near the physical pole at $k_0=\nu_+$, which is the proper (mean) frequency of the system, are related in \eqref{11} through a physical distribution function (positive definite for $k_0>0$). Therefore, such a relation would have the same form as the KB ansatz \eqref{2}.
A similar behaviour can also be seen in the other soluble model ($\delta m^{2} = 0$) discussed in the appendix. As will be shown in the next section, these features appear to be quite general in quenched models which are not necessarily soluble, as a consequence of causality.

\section{Generalized Kadanoff-Baym relation}\label{sec3}

In order to obtain a general relation between the correlated and spectral Green's
functions, which holds in the class of quenched models in \eqref{3} which are not necessarily exactly soluble, we consider the Feynman  Green's functions in the closed path formalism \cite{das}, which
have the 2 x 2 matrix form 
\begin{equation}
G=\left(
\begin{array}{ l c r }
  G_{++} & G_{+-}  \\
  G_{-+} & G_{--}  \\
\end{array}
\right)
\equiv
\left(
\begin{array}{ l c r }
  G_{++} & G_<  \\
  G_> & G_{--}  \\
\end{array}
\right),
\label{12}
\end{equation}
where the functions $G_>$ and $G_< $ are defined as the thermal averages 
\begin{equation}
iG_>(x,y)=\langle\phi(x)\phi(y)\rangle,\quad iG_<(x,y)=\langle\phi(y)\phi(x)\rangle.\label{13}
\end{equation}
Similarly, the corresponding self-energy functions have the 2 x 2 matrix structure
\begin{equation}
\Sigma =\left(
\begin{array}{ l c r }
  \Sigma_{++} & \Sigma_{+-}  \\
  \Sigma_{-+} & \Sigma_{--}  \\
\end{array}
\right)
\equiv
\left(
\begin{array}{ l c r }
  \Sigma_{++} & \Sigma_<  \\
  \Sigma_> & \Sigma_{--}  \\
\end{array}
\right).\label{14}
\end{equation}

It is more convenient to work with matrices of the form $\overline{M}=\sigma_3 M$ (where $\sigma_3$ is the third Pauli matrix) which lead to the simple matrix multiplication rules along the two branches of the closed time path contour \cite{spicka}. With this redefinition, the Green's functions and the self-energies  are connected by the Dyson equation
\begin{equation}
\overline{G}^{-1}(x,y)=\overline{G}^{(0)-1}(x,y)-\overline{\Sigma}(x,y),\label{15}
\end{equation}
where $G^{(0)}$ denotes the tree level Green's function. Using this relation together with the fact that the determinant of $\overline{G} $ equals to $G_R G_A$, 
one arrives at the Dyson-Keldysh equation
\begin{equation}
G_\gtrless=G_R(G_{R}^{(0)-1}G_{\gtrless}^{(0)}G_{A}^{(0)-1}+\Sigma_\gtrless)G_A.\label{16}
\end{equation}
 
In the general quenched model given in \eqref{3}, the exact Green's functions $G_R$ and $G_A$ have the  (Lippmann-Schwinger) form 
\begin{align}
G_R & =(1-G_{R}^{(0)}\Sigma_R)^{-1}G_{R}^{(0)},\notag\\
G_A & =G_{A}^{(0)}(1-\Sigma_AG_{A}^{(0)})^{-1},
\label{20}
\end{align} 
where the self-energy functions are given by
\begin{align}
\Sigma_R(x^0,y^0) & =\Sigma_A(x^0,y^0) = \Pi(x^0)\delta(x^0-y^0),\notag\\
\Sigma_\gtrless & = 0.\label{22}
\end{align}
We note, from \eqref{1} and \eqref{13}, that the Green's functions $G_\gtrless$, $G_{c}$ and $G_\rho$ can be related as
\begin{align}
G_{c}(x,y) & =G_>(x,y)+G_<(x,y),\notag\\
G_\rho(x,y) & =G_>(x,y)-G_<(x,y).\label{17}
\end{align}

Using the eqs. \eqref{16}, \eqref{20} and \eqref{22}, we obtain from \eqref{17} the following basic relation
\begin{equation}
G_{c,\rho}=(1-G_{R}^{(0)}\Sigma_R)^{-1}G_{c,\rho}^{(0)}(1-\Sigma_AG_{A}^{(0)})^{-1},\label{23}
\end{equation}
where the free correlated and spectral Green's function  have the forms
\begin{align}
G_{c}^{(0)}(x^0,y^0,\omega) & =-\frac{i}{\omega}\coth\left(\frac{\beta\omega}{2}\right)\cos\omega(x^0-y^0),\label{24}\\*
G_{\rho}^{(0)}(x^0,y^0,\omega) & =-\frac{1}{\omega}\sin\omega(x^0-y^0).\label{25}
\end{align} 

It follows from \eqref{24} and \eqref{25} that 
\begin{equation}
i\omega\coth\left(\frac{\beta\omega}{2}\right)
G_{\rho}^{(0)}(x^0,y^0,\omega)=-\frac{\partial}{\partial x^0}G_{c}^{(0)}(x^0,y^0,\omega),\label{26}
\end{equation}
 which is the fluctuation-dissipation theorem in coordinate space. Using this in \eqref{23} we obtain 
\begin{widetext} 
\begin{equation}
i\omega \coth\left(\frac{\beta\omega}{2}\right)
G_\rho(x^0,y^0)=
-(1-G_{R}^{(0)}\Sigma_R)^{-1}(x^0,z^0)\frac{\partial G_{c}^{(0)}(z_0,z_0')}{\partial z^0}
(1-\Sigma_A G_{A}^{(0)})^{-1}(z_0',y^0),\label{27}
\end{equation}
\end{widetext}
where integration over intermediate coordinates is understood and we have indicated only the time coordinates in the Green's functions for simplicity. Finally, integrating by parts to the left in \eqref{27} and using \eqref{22}, it is straightforward to obtain the following relation between the spectral and the correlated functions  
\begin{widetext}
\begin{equation}
i\omega\coth\left(\frac{\beta\omega}{2}\right)
G_\rho(x^0,y^0)=-\frac{\partial}{\partial x^0}G_{c}(x^0,y^0)+G_R(x^0,z^0)\frac{d\Pi (z^0)}{dz^0}G_{c}(z^0,y^0).\label{28a}
\end{equation}
\end{widetext}
Alternatively, replacing $\partial G_{c}^{(0)}/\partial z_0$ on the right hand side of \eqref{27} by $-\partial G_{c}^{(0)}/\partial z^0{}'$ and integrating by parts to the right, one obtains
\begin{widetext}
\begin{equation}
i\omega\coth\left(\frac{\beta\omega}{2}\right)
G_\rho(x^0,y^0)=\frac{\partial}{\partial y^0}
G_c(x^0,y^0)-G_c(x^0,z^0)\frac{d\Pi (z^0)}{dz^0}G_A(z^0,y^0).\label{28b}
\end{equation}
\end{widetext}
{ The two relations \eqref{28a} and \eqref{28b} are equivalent, which follows from the equality $G_R(x^0,y^0)=G_A(y^0,x^0)$ as well as from the fact that $G_c(x^0,y^0)$ and $G_ \rho(x^0,y^0)$ are respectively even and odd functions under the interchange $x^0\leftrightarrow y^0$.}

Equations \eqref{28a} and \eqref{28b}, which hold at all times, are one of our main results and some aspects of these relations are worth noting here. We remark that for a system which is always in thermal equilibrium,  $\Pi=0$, so that the last term in \eqref{28a} or \eqref{28b} vanishes at any time. In this case,  we would obtain the differential form of the fluctuation-dissipation theorem \eqref{26}, which in the Fourier transformed space has the form (see \eqref{1a}) 
\begin{align}
\widetilde{G}_c^{eq}(k_0,\omega) & =
-2\pi i\coth\left(\frac{\beta\omega}{2}\right)
\delta(k_0^2-\omega^2)\notag\\
& =\coth\left(\frac{\beta k_0}{2}\right)\widetilde{G}_\rho^{eq}(k_0,\omega).\label{29}
\end{align}
This relation exhibits a sharply peaked pole term at the natural frequency $k_0=\omega$ of the system in thermal equilibrium.

We also note that even out of equilibrium, the last term in \eqref{28a} does not contribute for $x^0<0$. This arises because the quench $\Pi(z^0)$ is non-zero only when $z^0\geq 0$ and the retarded Green's function $G_R(x^0,z^0)$ vanishes for $z^0>x^0$ by causality. Then, \eqref{28a} reduces to the form 
\begin{equation}
i\omega\coth\left(\frac{\beta\omega}{2}\right)
G_\rho(x^0,y^0)=-\frac{\partial}{\partial x^0}G_{c}(x^0,y^0).
\label{30a}
\end{equation}
Similarly, when $y^0< 0$, the last term in \eqref{28b} vanishes in which case one gets 
\begin{equation}
i\omega\coth\left(\frac{\beta\omega}{2}\right)
G_\rho(x^0,y^0)=\frac{\partial}{\partial y^0}
G_{c}(x^0,y^0).
\label{30b}
\end{equation}  
In the respective time domains, these have the forms similar to the fluctuation-dissipation theorem in equilibrium (although quantitatively they are very different) and we have explicitly verified these results in the soluble models given in \eqref{4}.  

 We now examine the Fourier transforms of eqs. \eqref{28a} and \eqref{28b} with respect to the time difference 
$t =x^0-y^0$. As we have pointed out following equation \eqref{10.1}, 
the $t$-integration is bounded in the range $|t|< 2|T|$ when $x^0$ and $y^0$ have the same sign  and leads to finite contributions in the $k_0$ space. On the other hand, the range of the $t$-integration is unbounded, $2|T|< |t|<\infty$ when $x^0$ and $y^0$                        
have opposite signs and yields  singular contributions. Such pole  terms occur in the Fourier transforms of \eqref{30a} and \eqref{30b} when ($x^0<0,\, y^0>0)$ and $(y^0<0,\, x^0>0)$, respectively. 
These differential forms would then lead, in the $k_0$ space, to the KB relation of the form in \eqref{2} between  the pole terms of $ \widetilde{G} _c (k_0,T)  $  and $\widetilde{G}_\rho(k_0,T)$ (see, for example, eqs. \eqref{10}, \eqref{9} and \eqref{11}).

\section{Non-relativistic generalized KB ansatz}\label{sec4}

Let us consider the non-relativistic limit of the theory in \eqref{3}, and denote by $\psi(x)$ the positive 
frequency part of the field (which would annihilate a particle in the free theory) and by  $\psi^{\dagger}(x)$     
its hermitian conjugate (which would create a particle in the free theory). In this case, it is useful to introduce the following thermal averages \cite{daniele}
\begin{equation}
ig_>(x,y)=\langle\psi(x)\psi^\dagger (y)\rangle,\quad ig_<(x,y)=\langle\psi^\dagger(y)\psi (x)\rangle.\label{31}
\end{equation}
Using \eqref{31}, one can then define the corresponding retarded/advanced Green's functions as
\begin{align}
G_R(x,y) & =\theta(x^0-y^0)\left[g_>(x,y)-g_<(x,y)\right],\notag\\
G_A(x,y) & =G_R^*(y,x).\label{32}
\end{align}

In our model, the exact non-relativistic retarded function in mixed space satisfies the equation
\begin{equation}
\left[i\frac{\partial}{\partial x^0}-E-\frac{\Pi(x^0)}{m}\right]G_R(x^0,y^0,E)=\delta(x^0-y^0),\label{33}
\end{equation}
where  $E =\mathbf{k}^2/2m$ and $\Pi(x^0)$ is the { general} quench described in \eqref{3} which is applied at non-negative times. The
solution of the above equation is easily determined to be 
\begin{align}
\lefteqn{G_R(x_0,y_0,E)}\notag\\
& =-i\theta(x^0-y^0)\exp\left(-i\int_{y^0}^{x^0}d\tau[E+\Pi(\tau)/m]\right).\label{34}
\end{align}

An important feature which follows from \eqref{34} is that the retarded Green's function satisfies the semigroup property
\begin{equation}
G_R(x_0,z_0,E)=iG_R(x_0,y_0,E)G_R(y_0,z_0,E),\label{35}
\end{equation}
for $x^0>y^0>z^0$. (It is important to note that such a property does not hold in the relativistic theory because of pair creation and annihilation processes). Using a procedure similar to that employed in the derivation of \eqref{16}, one can obtain the Dyson-Keldysh equation for the nonrelativistic theory which has the form
\begin{equation}
g_{\gtrless}=G_R [G_{R}^{(0)-1}g_{\gtrless}^{(0)}G_{A}^{(0)-1}+\Sigma_{\gtrless}]G_A,\label{36}
\end{equation}
where we have suppressed the arguments as well as the integration over intermediate coordinates for simplicity.

Apart from the factorization (semigroup) property \eqref{35}, another important ingredient in the derivation of
 a generalized KB ansatz in the non-relativistic theory is the proposal 
that the term inside the square bracket of \eqref{36} might be nearly diagonal in its time coordinates \cite{spicka}. This may be expected since the first term $G_{R}^{(0)-1}g_{\gtrless}^{(0)}G_{A}^{(0)-1}(z^0,z^0{}')$ contains delta functions which are strongly peaked at $z^0= z^0{}'$. Assuming that the non-diagonal    
contributions from this term as well as those in $\Sigma_{\gtrless}(z^0,z^0{}')$ may be neglected, one can derive the generalized KB ansatz \cite{lipavski,spicka}
\begin{align}
g_\gtrless(x^0,y^0) & =iG_R(x^0,y^0)g_\gtrless(y^0,y^0)\notag\\
&\quad -ig_\gtrless(x^0,x^0)G_A(x^0,y^0).\label{37}
\end{align}

We will next examine the validity of this causal ansatz, in the context of our exactly soluble model where $\Pi(x^0)$         has the form given in \eqref{4}. { Namely,} we will compare \eqref{37}
with the exact result \eqref{36} when $\Sigma_\gtrless =0$ and 
\begin{align}
g_{<}^{(0)}(x^0,y^0) & =-i N(E)\, e^{-iE(x^0-y^0)},\notag\\
g_{>}^{(0)}(x^0,y^0) & =-i(1+ N(E))\,e^{-iE(x^0-y^0)},\label{38}
\end{align}
where $N(E)$ denotes the equilibrium Bose-Einstein distribution function. Using \eqref{20} and \eqref{22}, it is convenient to write \eqref{36} in the form  
\begin{widetext}
\begin{align}
g_\gtrless(x^0,y^0) & = (1-G_{R}^{(0)}\Sigma_R)^{-1}(x^0,z^0)
g_{\gtrless}^{(0)}(z^0,z^0{}')(1-\Sigma_AG_{A}^{(0)})^{-1}(z^0{}',y^0)\nonumber \\
& = (1+G_{R}\Sigma_R)(x^0,z^0)g_{\gtrless }^{(0)}(z^0,z^0{}')(1+\Sigma_AG_{A})(z^0{}',y^0),\label{39}
\end{align}
\end{widetext}
where, in the last line, we have used \eqref{20} to identify
\begin{equation}
1 + G_{R}\Sigma_{R} = 1 + (1-G_{R}^{(0)}\Sigma_{R})^{-1} G_{R}^{(0)}\Sigma_R = (1 - G_{R}^{(0)}\Sigma_{R})^{-1},
\end{equation}
and so on. We can now substitute the exact expression for $G_R$ given in \eqref{34} and $G_{A}$ following from \eqref{32} into the relation \eqref{39}.

Let us first  set $\delta m^2=0$ in \eqref{4}, in which case we get, for example, the exact expression for $g_<(x^0,y^0)$ to be 
\begin{widetext}
\begin{equation}
g_< (x^0,y^0)=-i N(E)\, e^{-iE(x^0-y^0)}\left[1-i\frac{\Delta m}{m}\theta(x^0)e^{-i\frac{\Delta m}{2m}}\right]
\left[1+i\frac{\Delta m}{m}\theta(y^0)e^{i\frac{\Delta m}{2m}}\right].\label{40}
\end{equation}
\end{widetext} 
Using \eqref{40}, the approximate result in \eqref{37}  may be written (for $x^0>y^0$) as
\begin{align}
g_< (x^0,y^0) & =-iN(E)\, e^{-iE(x^0-y^0)}e^{-i\frac{\Delta m}{m}\theta(x^0)\theta(-y^0)}\notag\\
&\quad \times\left[1+\left(\frac{\Delta m}{m}\right)^2\theta(y^0)\right].\label{41}
\end{align}
We see that the expressions \eqref{40} and \eqref{41} are rather different. This implies that the assumption of neglecting non-diagonal terms made in \eqref{37} may not be justified in the presence of sharply peaked quenches.

On the other hand, for a regular quench of the form given in \eqref{4} with $\Delta m=0$, we find a
complete agreement between the expressions given in eqs. \eqref{36} and \eqref{37}. It turns out that in this model, $ig_<(x^0,x^0)= N(E)$ and  $ig_>(x^0,x^0)=1+N( E)$, so that these equations lead to the same result, namely
\begin{align}
g_<(x^0,y^0) & = N(E)[G_R(x^0,y^0)-G_A(x^0,y^0)],\notag\\
g_>(x^0,y^0) & =e^{\beta E}g_<(x^0,y^0).\label{42}
\end{align}
 The exact form of the retarded Green's function is given by (recall that $G_{A}(x^0,y^0) = G_{R}^{*}(y^0,x^0)$)
\begin{widetext}
\begin{equation}
G_R(x^0,y^0)=-i\theta(x^0-y^0)\exp\left(-i\left[\left(E+\frac{\delta m^2}{m}\theta(x^0)\right)x^0-\left(E+\frac{\delta m^2}{m}\theta(y^0)\right)y^0\right]\right).\label{43}
\end{equation}
\end{widetext}
Using \eqref{17} and \eqref{42}, one can also show that
\begin{equation}
G_{c}(x^0,y^0)=\coth\left(\frac{\beta E}{2}\right)G_\rho(x^0,y^0).\label{44}
\end{equation}                        

Although a simple KB relation holds in this  case, the model is useful to test the validity of certain approximations made in realistic theories which are not exactly soluble. A relevant issue   of the generalized KB ansatz \eqref{37} concerns the role of the involved unknown retarded and advanced propagators. 
An  approximation which is  much used in practice is based on the assumption that the pole term, which involves a delta function in Fourier space, would give the main contribution. Let us check this assumption in our soluble model, where the Fourier transform of the exact retarded propagator \eqref{43} has a sharp peak at the quasiparticle energy. We get 
\begin{widetext}
\begin{equation}
\widetilde{G}_R(k_0,T) = \theta(T)\left[
\frac{1-e^{2i(k_0-E-\frac{\delta m^2}{m})T}}{k_0-E-\frac{\delta m^2}{m}}+\frac{e^{2i(k_0-E-\frac{\delta m^2}{m})T}}{k_0+i\epsilon-E-\frac{\delta m^2}{2m}}
\right] + \theta(-T)\left[
\frac{1-e^{-2i(k_0-E)T}}{k_0-E}+\frac{e^{-2i(k_0-E)T}}{k_0+i\epsilon-E-\frac{\delta m^2}{2m}}\right].\label{45}
\end{equation}
\end{widetext}

This result exhibits a pole at $k_0=E+\delta m^2/2m-i\epsilon$, which leads to a delta function contribution 
in the spectral  function  $\widetilde{G}_\rho=2i\,\text{Im}\, \widetilde{G}_R$. Going back to the mixed space, this term yields 
the following form  for the approximate retarded Green's function
\begin{equation}
G_R^q (x^0,y^0)=-i\theta(x^0-y^0)e^{-i E(x^0-y^0)}e^{-i\frac{\delta m^2x^0}{m}}. 
\label{46}
\end{equation}
This form agrees with \eqref{43} when $x^0>0$ and $y^0<0$. This occurs because the pole terms in $\widetilde{G}_R$ come precisely from this region, as we have pointed out earlier. On the other hand, \eqref{46} is rather different from the exact result \eqref{43} when $x^0<0$ or $y^0>0$, which shows that the quasiparticle approximation is not satisfactory in general.

\section{Summary}\label{sec5}
  We have derived, ~in the context of the class ~of non-equilibrium ~quenched models \eqref{3}, a ~generalization of the Kadanoff-Baym ansatz. This extension ~involves ~a generalized differential form relating the exact spectral and correlated Green's functions, which holds at all times (see eqs. \eqref{28a} and \eqref{28b})
\begin{align}
i\omega\coth\left(\frac{\beta\omega}{2}\right)G_\rho(x^0,y^0,\omega) & =
-
\frac{\partial}{\partial x^0}
G_{c}(x^0,y^0,\omega)+\notag\\
&\quad (G_R\dot{\Pi}G_{c})(x^0,y^0,\omega),\label{47}
\end{align}
where $\omega$ is the energy given by \eqref{8} and the dot denotes the derivative of the quench $\Pi(z^0)$   with respect to time.
This equation can be explicitly verified in the exactly soluble models \eqref{4}. 
We note that in equilibrium $\dot{\Pi}$ vanishes, so that the above equation reduces to the differential form of the  fluctuation-dissipation theorem which relates $G_c$ and $G_\rho$ through a thermal factor. On the other hand, when the system is out of equilibrium $\dot{\Pi}$ is significant and owing to the presence of the retarded Green's function $G_R$ in \eqref{47}, $G_c$ and $G_\rho$ become in general independent functions. Nevertheless, even out of equilibrium, the last
term vanishes by causality when $x^0<0$, in which case \eqref{47} simplifies to a particular differential  equation \cite{britto}. In the Fourier space, this equation leads to pole terms of $\widetilde{G}_{c}(k_0,T,\omega)$ and  $\widetilde{G}_\rho(k_0,T,\omega)$ at the natural frequency of the system, which are related through the usual KB ansatz \eqref{2}. We may interpret  such a behavior from  a 
physical point of view by noting that in general the sudden quench takes the system out 
of equilibrium. However, when $k_0$ is close to the natural frequency of the system, this may remain in a state of near-equilibrium. 
   
We have also examined the non-relativistic limit of simple (quenched) models which are exactly soluble. These models provide 
a framework for testing certain assumptions made in realistic (but not exactly soluble) many-body theories,
concerning alternative generalizations of the Kadanoff-Baym ansatz. We have shown that the neglect of non-diagonal contributions
in the Dyson-Keldysh equation \eqref{36}, which is an important condition for the derivation of such generalizations, may not be justified in the presence of sharply peaked quenches. On the other hand, this procedure is valid for regular quenches and leads to the causal relation \eqref{37}.
However, we have verified that in this case, the quasiparticle approximation \eqref{46} for the 
retarded propagator is consistent with the exact result \eqref{43} only in a particular time sector.
Thus, we conclude that the quasiparticle ansatz may not be appropriate in general.

\appendix
\renewcommand{\theequation}{A.\arabic{equation}}
\setcounter{equation}{0}  
\subsection*{Acknowledgments}
A. L. B. and J. F. would like to thank Conselho Nacional de Pesquisa (CNPq), Brazil, for financial support.
\section*{Appendix}
In this appendix we discuss the exactly soluble model obtained from  \eqref{3} and \eqref{4}  by setting $\delta m^2=0$.  In mixed space, the exact retarded and correlated Green's functions, for this model, have the forms (we indicate only their time dependence for simplicity)
\begin{widetext}
\begin{align}
G_R(x^0,y^0) & = G_{R}^{(0)}(x^0,y^0)
+(\Delta m)G_{R}^{(0)}(x^0,0)G_{R}^{(0)}(0,y^0), \label{a1}\\
G_{c}(x^0,y^0) & = G_{c}^{(0)}(x^0,y^0) + (\Delta m)[G_{R}^{(0)}(x^0,0)G_{c}^{(0)}(0,y^0)+G_{R}^{(0)}(y^0,0)G_{c}^{(0)}(0,x^0)]\notag\\
&\qquad + (\Delta m)^2G_{R}^{(0)}(x^0,0)G_{R}^{(0)}(y^0,0)G_{c}^{(0)}(0,0),\label{a2}
\end{align}
\end{widetext}
where $G_{c}^{(0)}$ is given in \eqref{24} and $G_{R}^{(0)}$ has the form
\begin{align}
\lefteqn{G_{R}^{(0)}(x^0,y^0) =\int\frac{dk_0}{2\pi}e^{-ik_0(x^0-y^0)}\frac{1}{(k_0+i\epsilon)^2-\omega^2}}\notag\\
&\qquad = -\frac{\theta(x^0-y^0)}{\omega}e^{-\epsilon(x^0-y^0)}\sin\omega(x^0-y^0).\label{a3}
\end{align}

The infinitesimal Feynman parameter $\epsilon$ in \eqref{a3} is to be taken to zero only at the end of the calculation. When the time difference $x^0-y^0$ is finite, the regularizing exponential factor may be set equal to unity, in which case one gets back the usual expression for the free retarded Green's function. However, we will keep this factor for generality. Let us now look, for example, at the last term in \eqref{a2} which may be written as
\begin{widetext}
\begin{equation}
\left(\Delta m\right)^2 G_{R}^{(0)}(x^0,0)G_{R}^{(0)}(y^0,0)G_{c}^{(0)}(0,0) 
=\frac{\theta(x^0)\theta(y^0)}{i\omega}
\left(\frac{\Delta m}{\omega}\right)^2 \coth\left(\frac{\beta\omega}{2}\right)
e^{-\epsilon(x^0+y^0)}
\sin(\omega x^0)\sin(\omega y^0).\label{a4}
\end{equation}
\end{widetext}

For large times (after the quench) the exponential factor suppresses the  contributions 
from the rapidly oscillatory trigonometric functions. This is a general feature in quenched models, where the Feynman parameter $\epsilon$  plays the role of the inverse relaxation time \cite{frenkel2}.
It is easy to verify that, for $x^0 <0$ and for $y^0<0$, the exact Green's functions in \eqref{a1} and \eqref{a2} satisfy respectively the relations
\begin{subequations}
\begin{align}
& i\omega\coth\left(\frac{\beta\omega}{2}\right)
G_\rho(x^0,y^0) =-\frac{\partial}{\partial x^0}G_{c}(x^0,y^0),\label{a5a}\\
& i\omega\coth\left(\frac{\beta\omega}{2}\right)
G_\rho(x^0,y^0)=\frac{\partial}{\partial y^0}G_{c}(x^0,y^0).\label{a5b}
\end{align}
\label{a5}\end{subequations}

Taking the Fourier transforms of \eqref{a1} and \eqref{a2} with respect to the time difference $t=x^0-y^0$ and using the forms of the free Green's functions, we obtain 
\begin{widetext}
\begin{align}
\text{Im}\widetilde{G}_R(k_0,T) & =\widetilde{G}_\rho(k_0,T)/2i\notag\\
&=-\pi\text{sign}(k_0)\delta(k_0^2-\omega^2)
+ \frac{\Delta m}{2\omega^2}
\left[
\cos(2\omega T)\frac{\cos 2k_0T}{k_0}-
\frac{1}{2}\left(\frac{\cos2(k_0+\omega)T}{k_0+\omega}+\frac{\cos2(k_0-\omega)T}{k_0-\omega}\right)
\right],\label{a6}\\
\widetilde{G}_{c}(k_0,T) & =
-2i\coth\left(\frac{\beta\omega}{2}
\right)
\left\{\frac{}{} \pi\delta (k_0^2-\omega^2)\right.\nonumber \\
&\quad - \frac{\Delta m}{2\omega^2}
\left[\pi\sin(2\omega T)\delta(k_0)+\sin(2\omega T)\frac{\sin 2k_0 T}{k_0}+\frac{1}{2}
\left(
\frac{\cos2(k_0+\omega)T}{k_0+\omega}-\frac{\cos2(k_0-\omega)T}{k_0-\omega}
\right)\right]\nonumber\\
&\quad +\theta(T)\left(\frac{\Delta m}{\omega}\right)^2\frac{1}{2\omega}
\left[
\frac{\sin2(k_0+\omega)T}{k_0+\omega}+\left.\frac{\sin 2(k_0-\omega)T}{k_0-\omega}-2\cos(2\omega T)\frac{\sin(2k_0 T)}{k_0}
\right]\right\}.\label{a7}
\end{align}                                                                                                                                                    
\end{widetext}

These results show that near the physical pole $k_0=\omega$,  the exact correlated and spectral Green's functions, $\widetilde{G}_{c}$ and $\widetilde{G}_\rho$, are simply related  as
\begin{equation}
\widetilde{G}_{c}(k_0,T)\simeq\coth(\beta k_0/2)\widetilde{G}_\rho(k_0,T).\label{a8}
\end{equation}              
As we have mentioned, in the mixed space these poles come from the regions where $x^0$ and $y^0$ have opposite signs, so that the range of the difference $t=  x^0-y^0$ is unbounded. These 
regions [($x^0<0$, $y^0>0$) and ($y^0<0$, $x^0>0$)] correspond respectively to the domain of validity of the differential forms \eqref{a5a} and \eqref{a5b}, which lead
 to the KB relation \eqref{a8}.

\end{document}